\documentclass[%
 aip,
 jmp,%
 amsmath,amssymb,
 reprint,%
nofootinbib]{revtex4-1}
\usepackage{graphicx}% Include figure files
\usepackage{dcolumn}% Align table columns on decimal point
\usepackage{bm}% bold math
\usepackage{subfigure}
\usepackage{color} 

\begin{document}

\preprint{AIP/123-QED}

\title[]{A New Scheme for High-Intensity Laser-Driven Electron Acceleration in a Plasma}
%\thanks{Here we consider the spatially unbounded plasma-bunch systems which can also be referred to as the spatially unbounded  plasma-beam systems.}

\author{A. A. Rukhadze}
\email{rukh@fpl.gpi.ru}
%\altaffiliation{A. A. Rukhadze and S. P. Sadykova contributed equally to this work.}
\affiliation{Prokhorov General Physics Institute,
Russian Academy of Sciences, Vavilov Str., 38., Moscow, 119991, Russia}
\author{S. P. Sadykova}%
\altaffiliation{Electronic mail: \textcolor{blue}{Corresponding author - s.sadykova@fz-juelich.de}}
%\altaffiliation{A. A. Rukhadze and S. P. Sadykova contributed equally to this work.}
 %\email{Corresponding author, saltanat@physik.hu-berlin.de.}
\affiliation{Forschungszentrum Julich, J{\"u}lich Supercomputing Center, J{\"u}lich, Germany%\\This line break forced with \textbackslash\textbackslash
}%
\author{T. G. Samkharadze}
%\affiliation{Prokhorov General Physics Institute,
%Russian Academy of Sciences, Vavilov Str., 38., Moscow, 119991, Russia}
%\author{S.P. Andreev}
%\affiliation{Prokhorov General Physics Institute,
%Russian Academy of Sciences, Vavilov Str., 38., Moscow, 119991, Russia}
\author{P. Gibbon}
\affiliation{Forschungszentrum Julich, J{\"u}lich Supercomputing Center, J{\"u}lich, Germany}
% \homepage{http://www.Second.institution.edu/~Charlie.Author.}
%\affiliation{%
%Second institution and/or address%\\This line break forced% with \\
%}%

\date{\today}% It is always \today, today,

\begin{abstract}
We propose a new approach to high-intensity laser-driven electron acceleration in a plasma. Here, we demonstrate that a plasma wave generated by a stimulated forward-scattering of an incident laser pulse can be in a longest acceleration phase with an incident laser wave. This is  why the plasma wave has the maximum amplification coefficient which is determined by the breakdown (overturn) electric field in which the  acceleration of injected relativistic beam electrons occurs. We estimate qualitatively the acceleration parameters of relativistic electrons in  the field of a plasma wave generated at the stimulated forward scattering of a high-intensity laser pulse in a plasma.

%Valid PACS numbers may be entered using the \verb+\pacs{#1}+ command.
\end{abstract}

%52.40.Mj; 52.35.-g; 41.60.-m.}% PACS, the Physics and Astronomy
%Use showkeys class option if keyword
\maketitle

%\section{\label{sec:Int}Introduction}

\begin{figure}
\centering
\includegraphics[width=0.7\linewidth]{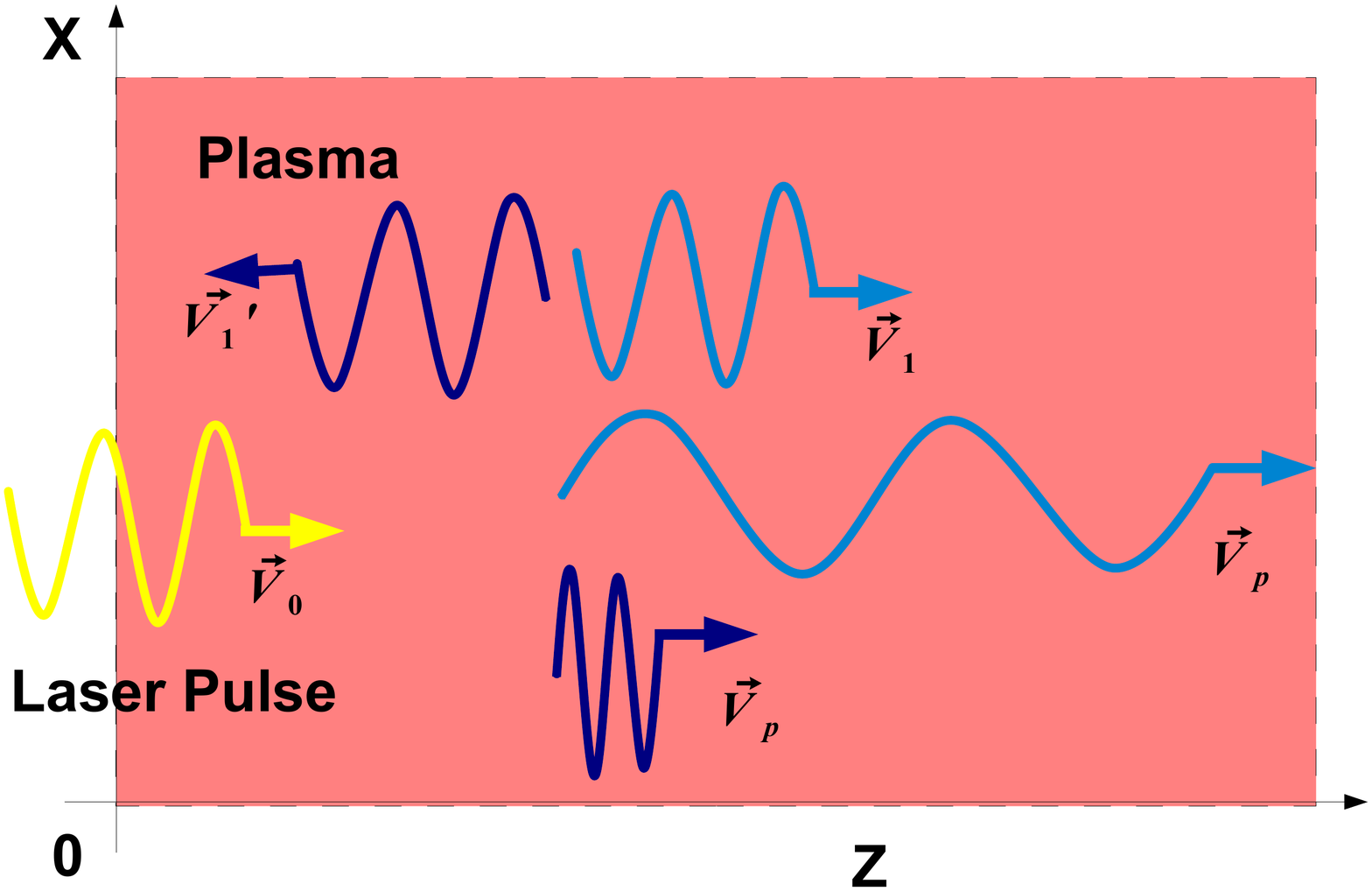}
\caption{Schematic 2D-illustration of interaction of the high-intensity laser pulse with plasma which generates the plasma wave and forward- and backward-scattered waves with $\vert \vec {V_p}\vert=\vert\vec {V_0}\vert=\vert\vec {V_1}\vert$, $k_0\simeq k_1>k_p$  (forward-scattered wave), $V_1'$ - back-scattered wave, where $\vert k_0\vert \simeq \vert -k_1\vert<\vert k_p \vert$.}
\label{Fig:1}
\end{figure}

\begin{figure}
\centering
\includegraphics[width=0.7\linewidth]{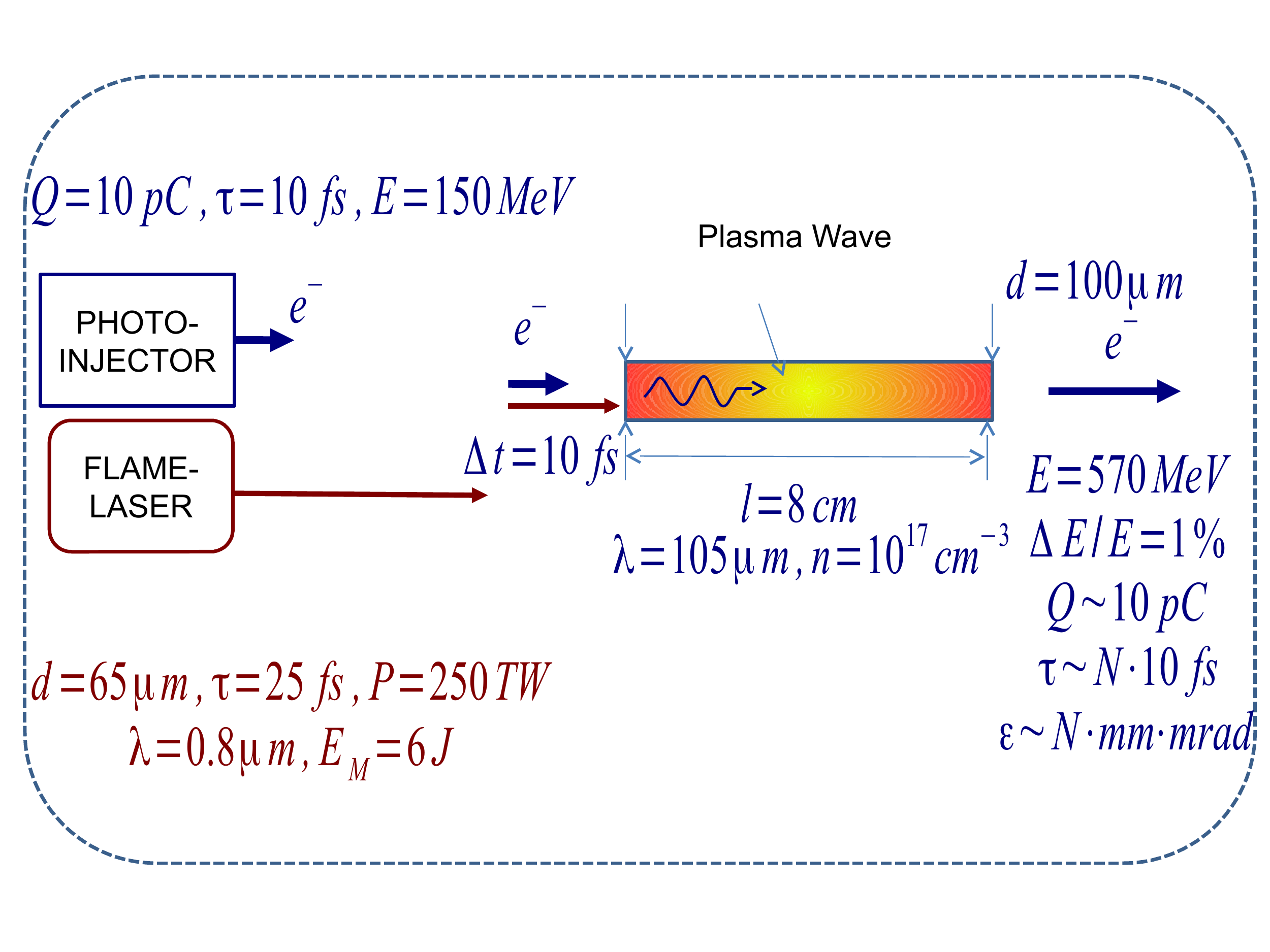}
%\vspace{-190pt}
\caption{Schematic drawing of the experiment made at the \texttt{SPARC\char`_LAB} facility of INFN-LNF in Frascati, Italy \cite{Ross}.}
\label{Fig:2}
\end{figure}

 During the past few decades plasma accelerators have attracted increasing interest of scientists from all over the world due to its compactness, much cheaper construction costs compared to those for conventional one and various applications ranging from  high energy physics to medical and industrial applications. An intense electromagnetic pulse can create a plasma oscillations through  the stimulated scattering. Electrons trapped in the plasma wave can be accelerated to high energy.\\
\indent  The idea to accelerate the charged particles in a plasma medium using collective plasma wave fields generated by the high-energy electron beams belongs to the Soviet physicists G. I. Budker, V. I.  Veksler and Ia. B. Fainberg \cite{Budker,Veksler,Fainberg} in 1956, whereas assumptions for  generation of plasma  Langmuir waves by nonrelativistic electron bunches propagating through plasma were first made earlier in 1949 \cite{1, 2}.  High-energy bunch electrons generate a plasma wave in such a way that the energy from a bunch of electrons is transferred to the plasma wave through stimulated Cherenkov resonance radiation producing high acceleration electric fields. Later on, another acceleration scheme using a laser \cite{Dawson2} or time-shifted sequence of  bunched high-energy electrons injected into a cold plasma was proposed \cite{Dawson}.  In recent experiments at the Stanford Linear Accelerator Center it was shown that an energy gain of more  than 42 GeV was achieved in a meter long plasma wakefield accelerator, driven by a 42 GeV electron beam \cite{Blum}. For a detailed review about the modern status of this  research  field we would like to refer a reader to \cite{Joshi, Liver}.  \\
\indent However, the experiments of the 60s and 70s demonstrated that efficiency of acceleration using the high-energy beams is much less than the expected one and the  generated field is much lower than a breakdown (overturn) electric field \cite{Akh}:

\begin{equation}\label{4}
{E_p}_{max}=2 \pi m V_p\omega_p /e.
\end{equation}
where $e$- electron charge, $m$ - its mass, $V_p$ - plasma wave phase velocity, $V_p=\omega_p/k_p \simeq c$ where $k_p$ - plasma wave vector, $\omega_p$ - plasma frequency, ${\omega_{p}}=\sqrt{4\pi e^2 n_{e}/m}$ with $n_e$ being the electron density and m - its mass. The explanation for the experiments failure was given in the work \cite{9} where it was shown that trapping of electrons in a generated by a beam plasma wave occurs and as a result growth of the field amplitude stops when the field amplitude is much less than that of the breakdown field \cite{4, 5}.\\
\indent With the apperance of the high-intensity lasers in the 80s, a new era of plasma acceleration has begun. The hopes were feeded by  possibility of the stimulated scattering of a laser pulse by plasma electrons in a rare plasma with the generation of high-intensity  longitudinal plasma wave where the electron trapping does not occur and, as a result, high fields with respect to the breakdown field  ${E_p}_{max}$ can be generated. However, there are still a lot of unsolved problems related to the development of instabilities hindering the laser-driven plasma-based acceleration \cite{Liver}.   \\
	\indent We presume that the key issue of the laser-driven plasma-based acceleration is the following: inspite of the fact that the  instability induced by the backward (backwards the laser pulse) scattering generating a plasma wave or a wake has a maximum increment $\delta_1$ compared to that generated by a forward (towards the injection of a laser pulse) one $\delta_2$ by $\sim 2\omega_0/\omega_p $ times ($\omega_0$ - laser pulse frequency) \cite{4} (see Fig. \ref{Fig:1}), this acceleration scheme is not suitable for particle acceleration because such lasers have a very short laser pulse length. Since the wave vector of a plasma wave $\vec k_p$ is equal to the double magnitude of that of a laser pulse ($k_p \simeq 2\omega_0/c$), the phase velocity of a plasma wave is quite low. Due to this fact the wave leaves behind both the laser and the back scattered waves getting localized at the back to the front of the laser pulse (this is why it is called as a wake). 
	As a result the plasma wave gets soon out of the acceleration phase with the laser wave and the electron beam injected into the plasma gets soon out of a phase with the plasma wave what halts the acceleration process. In the range of carried out experiments the acceleration effeciency estimates to not higher than 20-50 \%. Recently, the \texttt{SPARC\char`_LAB} facility of INFN-LNF in Frascati, Italy,  reported the observation of electrons acceleration with an energy gain of 420 MeV (300 \%) out of 150 MeV injected using a 250 TW laser system ($\lambda=800$ nm, $\tau=25$ fs, $E_M=6$ J) \cite{Ross} (see Fig. \ref{Fig:2}).\\
	\indent	In this work, we are first to propose another acceleration scheme, namely: a stimulated forward-scattering based plasma acceleration. In addition, on a base of such model we would like to give some explanations for the experiment \cite{Ross}. Due to the stimulated laser forward-scattering a plasma wave  is generated as well. In this case, the plasma wave and laser wave can stay in acceleration resonance for a much longer time.\\
%	\indent  In the present work this idea along with the employment of ultrarelativistic electron bunch is discussed at the qualitative  level. Namely, we make  an estimation of plasma parameters, maximum amplitude of the generated wakefield when the ultrarelativistic  electron and proton bunches are employed and plasma, bunch lengths at which the maximum amplitude of the wakefield can be gained. 	
%\section{Description of the Model. Waves Phase velocities}
Let the laser pulse with Langmuir frequency ${\omega_{0}}$ be injected into the cold plasma at  $\boldsymbol{{\omega_0}^2>>{\omega_p}^2}$. Consider the case when  $Z||\vec{V_0}$, $\vec{V_0}$, $\vec{V_1}$ and $\vec{V_p}$ - phase velocities of laser, forward-scattered and plasma waves.
%\vspace{-15pt}
Here, we employ the CGS system of units. The parametric resonance for the stimulated forward-scattering can be written as following:
%The surface wave is a wave of \textbf{$E$-type} with the nonzero field components $E_x,E_z, B_y$, which satisfy the following system of equations \footnotemark[1]:
%\footnotetext[1]{\textcolor{blue}{A.F. Alexandrov, L.S. Bogdankevich, A.A. Rukhadze,  \textit{Principles of Plasma Electrodynamics} (Springer, Heidelberg, 1984)}}
%\begin{alertblock}{}
\begin{equation}\label{1}
\begin{gathered}
\omega_0=\omega_1+\omega_p \hfill \\
k_0=k_1+k_p,
\end{gathered}
\end{equation}
%\end{alertblock}
where $k_0$, $k_1$ and $k_p$ are the wave vectors of the incident laser pulse, forward-scattered wave and plasma wave respectively, $\omega_0$, $\omega_1$ - corresponding frequencies: 
\begin{equation}\label{2}
\begin{gathered}
\omega_0=\sqrt{\omega_p^2+k_0^2 c^2} \hfill \\
\omega_1=\sqrt{\omega_p^2+k_1^2 c^2},
\end{gathered}
\end{equation}
After having solved the system of Eqs. (\ref{1}) and (\ref{2}) one can find out that the phase velocities of three waves are equal
\begin{equation}
\label{3}
\left(V_0=\frac{\omega_0}{k_0}\right)=\left(V_1=\frac{\omega_1}{k_1}\right)=\left(V_p=\frac{\omega_p}{k_p}\right)=c\left(1+\frac{\omega_p^2}{2\omega_0^2}\right),
\end{equation}
where ${\omega_p}^2/{\omega_0}^2<<1$.
%\section{Breakdown electric field}
The condition for the resonance interaction (\ref{3}) can be satisfied for a sufficiently long time until the amplitude of the plasma  wave becomes higher than that of the laser incident wave and the reversed process of feeding back the incident wave starts. If the laser is powerful enough than the instability will keep growing until a breakdown (overturn) of the plasma wave occurs, i.e. when the magnitude of the  saturation plasma wave amplitude becomes equal to ${E_p}_{max}$ (see Eq. \ref{1}).  %\footnotemark[3]:
%\footnotetext[3]{\textcolor{blue}{A.I.Akhiezer, I.A.Akhiezer, R.V.Polovin, A.G.Sitenko and K.N.Stepanov, \textit{Plasma Electrodynamics}, V.1. Linear Theory, V.2. Nonlinear theory and fluctuations, International Series of Monographs in Natural Philosophy, Vol.68, Vol.69 (Oxford-New York: Pergamon Press, 1975)}} 
 The acceleration constraint (\ref{4}) enables us to controle the acceleration process. In order to determine the electron trapping condition for acceleration we need to estimate the duration of acceleration, i.e. the time interval during which the phase velocities of a plasma wave and that of an electron remain approximately equal to each other. This condition possibly was taken into account during the \texttt{SPARC\char`_LAB} experiment in Frascatti with injected electron beam energies of 150 MeV. \\
	\indent Taking into account this condition we can estimate the duration of acceleraion of an ultrarelativistic electron with an initial energy $\varepsilon=mc^2 (\gamma-1)$, $\gamma=1/\sqrt{1-{u_0}^2/c^2}>>1$, $u_0$ is the speed of an electron beam. The speed of such an electron is the following: 

\begin{equation}
\label{5}
%\begin{gathered}
\frac{u_0}{c}=1- \frac{1}{2\gamma^2}
%\end{gathered}
\end{equation}

 Using the following estimation formula for acceleration time:

\begin{equation}
\label{6}
(V_p-u_0)\tau_f \approxeq c\pi/\omega_p.
\end{equation}

 and Eqs. (\ref{3}) and (\ref{5}) we can detemine the acceleration time :

\begin{equation}
\label{7}
\tau_f \approxeq \frac{2\pi\gamma^2\omega_0^2}{\omega_p(\omega_p^2\gamma^2+\omega_0^2).}
\end{equation}
Taking into account Eq. (\ref{4}) and (\ref{7}) the following momentum and energy growth can be obtained:
%\begin{alertblock}{} 
\begin{equation}\label{8}
\Delta P\approx e {E_p}_{max} \tau, \:\:\:\:\: \Delta\varepsilon \approx e {E_p}_{max} \tau c.
\end{equation}
\indent Let us make some estimations for a laser-driven plasma-based electron accelerator and compare with the results obtained at the 
\texttt{SPARC\char`_LAB} facility of INFN-LNF in Frascati, Italy.

\begin{itemize}
\item {\textbf{Estimated parameters}}
\item[-]  $\omega_0=2.35\cdot 10^{15}$ s$^{-1}$ ($\lambda=800 $ nm), $\omega_p=1.8\cdot 10^{13}$ s$^{-1}$, $n_e=10^{17}$ cm$^{-3}$
\item[-] An electron with energy growth of $150$ MeV ($\gamma=297$) can gain the maximum energy $\Delta\varepsilon\simeq 0.3\cdot 10^{12}$ eV in the  saturation field of ${E_p}_{max}\simeq 1.9 \cdot 10^{9}$ V/cm during $\tau_f=5\cdot 10^{-9}$ s over the maximum acceleration length of $L \simeq 150$ cm. Provided that the length of a capillary could be of the same size: $L \simeq 150$ cm, the corresponding energy growth for the \texttt{SPARC\char`_LAB} facility can be estimated to much higher energy growth of $\Delta\varepsilon\simeq 15 \cdot 10^{9} $ eV compared to the obtained one. 
\item[-] The acceleration time in a frame of  the stimulated back-scattered model can be determined as $\tau_b= \pi/\omega_p$, $\tau_b \simeq 10^{-13}$ s which is much less than that for the forward-scattered case: $\tau_f\simeq 10^{-9}$ s by approx. $\gamma^2$ times (provided that  $\gamma>> \omega_0/\omega_p$). The corresponding plasma wave lengths will be $\lambda_b= \pi c/\omega_0$, $\lambda_b\simeq 0.5 \mu  $m and $\lambda_f= 2\pi c/\omega_p$, $\lambda_f= 105 \mu$m.
\end{itemize}

\section{Results and Discussions}
\indent In the present work for the first time the analytical problem of interaction of a high-intensity laser pulse with  plasmas in a frame of a stimulated forward-scattering model  has been solved and a new approach to solution of the problem was proposed.\\
 \indent The acceleration scheme employing the stimulated backward-scattered wave for particle acceleration in a wakefield is not suitable for particle acceleration  because high-intensity lasers have a very short laser pulse length leading to a very short interaction time between the  injected electron beam and plasma wave. Instead, our new approach employing the stimulated  forward-scattering can provide more durable particle trapping time inside the field of a plasma wave of a much longer length compared o the back-scattered model where electrons just slip off the wave. In Eq. for a breakdown electric field for illustration purpose we took a constant topf field because the exact electric field profile during the instability growth is not known. However, the time interval during which the breakdown field can be gained is much less than the acceleration time what justifies our choice of the field profile.  For a better and detailed research we are planning to run more  extensive simulations of the considered phenomena using the \textbf{KARAT} code \cite{KAR}. \\
 \indent In order to acquire the maximum acceleration effeciency during the experiment one would need to make the scattered at various angles radiation except the forward-scattered one be absorbed by plasma surrounding surfaces similar to the scheme employed for a SHF (superhigh frequency) plasma amplifier.  

\begin{acknowledgments}
 S.P. Sadykova would like to express her gratitude to the Helmholtz Foundation for its financial support of the work.
\end{acknowledgments}

\appendix

\nocite{*}
%\bibliography{aipsamp}% Produces the bibliography via BibTeX.

\end{document}